\documentclass[11pt]{article}
\usepackage{hyperref}
\begin{document}
\title{Geometry of elastic hydrofracturing by injection of an over pressured non-Newtonian Fluid}
\author{Mariano Cerca, Jazmin Chavez Alvarez, Bernardino Barrientos,\\ Enrique Soto and Carlos Mares 
\\\\ \vspace{6pt}Universidad Nacional Aut\'onoma de M\'exico,\\ Blvd Juriquilla 3001, Juriquilla, Quer\'etaro, 76230, M\'exico\\  Centro de investigaciones en \'Optica,\\ Loma del Bosque \#115, Le\'on, 37150, M\'exico.}

\maketitle
\begin{abstract}

The nucleation and propagation of hydrofractures by injection of over pressured fluids in an elastic and isotropic medium are studied experimentally. Non-Newtonian fluids are injected inside a gelatine whose mechanical properties are assumed isotropic at the experimental strain rates. Linear elastic theory predicts that plastic deformation associated to breakage of gelatin bonds is limited to a small zone ahead of the tip of the propagating fracture and that propagation will be maintained while the fluid pressure exceeds the normal stress to the fracture walls (Ch\'avez-\'Alvarez,2008) (i.e., the minimum compressive stress), resulting in a single mode I fracture geometry.  However, we observed the propagation of fractures type II and III as well as nucleation of secondary fractures, with oblique to perpendicular trajectories with respect to the initial fracture. In the \href{http://ecommons.library.cornell.edu/handle/1813/14122}{Video} experimental evidence shows that the fracture shape depends on the viscoelastic properties of gelatine coupled with the strain rate achieved by fracture propagation.

\end{abstract}
\begin{enumerate}
\item M.J. Ch\'avez \'Alvarez, L.M. Cerca Mart\'inez, \emph{Analogue simulation of magama rheology during dike emplacement: A preliminary study based on field observations and rheological determinations of materials, Bolletino di Geofisica,\textbf{49}, 29-34 (2008).}

\end{enumerate}

\end{document}